\input epsf.tex

\documentstyle[prl,aps,twocolumn,floats]{revtex}

\begin{document}

\title{Mott transition, antiferromagnetism, and unconventional
superconductivity  in  layered
organic superconductors
}
\author{S. Lefebvre$^{1,2}$, P. Wzietek$^1$, S. Brown$^3$, C.
Bourbonnais$^2$, D. J\'erome$^1$, C.
M\'ezi\`ere$^4$, M. Fourmigu\'e$^4$  and P. Batail$^4$}
\address{ $^1$Laboratoire de Physique des Solides (CNRS, U.R.A.2),
B\^atiment 510 Universit\'e de
Paris-sud, 91405 Orsay, France}
\address{$^2$CERPEMA,
D\'epartement de Physique, Universit\'e de Sherbrooke, Sherbrooke,
Qu\'ebec, Canada J1K 2R1}
\address{$^3$Department of Physics and Astronomy, University of California
at LosAngeles}
\address{ Los
Angeles, CA0025, USA}
\address{$^4$Institut des Mat\'eriaux de Nantes, 44072 Nantes, France}

\author{
\widetext
\begin{center}
\begin{abstract}
\parbox{14cm}
{The  phase diagram of the layered organic superconductor
$\kappa$-(ET)$_{2}$Cu[N(CN)$_{2}$]Cl has been accurately measured   from a
combination
of $^{1}$H NMR and AC susceptibility techniques under helium gas pressure. The
domains of stability of antiferromagnetic and superconducting long-range
orders in the pressure
{\it vs}  temperature  plane have been  determined. Both phases  overlap
through  a first-order
boundary that separates two regions of inhomogeneous phase coexistence.
The boundary curve is found to merge
with another first order line related to  the
metal-insulator transition in the paramagnetic region. This transition is
found
to evolve into a crossover regime above a critical point at higher
temperature. The
whole phase diagram  features a   point-like region where metallic,
insulating, antiferromagnetic
and non s-wave superconducting phases all meet.   }
\end{abstract}
\end{center}
}
\maketitle

\narrowtext
The  determination of the  conditions giving rise to superconductivity (SC) in layered organic
conductors constitute one of the chief objectives in understanding  the physics of these strongly
correlated   electronic materials \cite{McKenzie97}. Closely bound to the now classical issue of proximity
of antiferromagnetism (AF) in the emergence of   superconductivity   
 stands the  problem of the `normal' phase which, 
depending on pressure conditions in these systems, is either a Mott insulator (MI) or an unconventional
metal \cite{Ito96}. A pressure-driven
metal-insulator transition can be thus revealing  of the  strong
coupling
conditions for electrons that are responsible for broken symmetry states
\cite{Kanoda97}.

In this
matter, the phase diagram of the series of layered organic
superconductors $\kappa
$-(BEDT-TTF)$_2$X as a function of both hydrostatic and chemical   (or anion X substitution) pressures  is
set to stand out of the debate.  By chemical means, the study of anion substituted compounds
 has allowed few discrete shifts of the pressure scale. Thus for X=~Cu[N(CN)$_2$]Br and X=~Cu(NCS)$_2$, 
 experiments adduce growing  evidence for an unconventional metal and  a non s-wave SC state
\cite{Kanoda97,Wzietek96},  
whereas AF order is shown to become in turn  
 stable  on the deuterated X= d$^n$-Cu[N(CN)$_2$]Br  compound
 \cite{Kanoda97,Miyagawa95}.

Among all members of the series $\kappa -$(BEDT-TTF)$_2$Cu[N(CN)$_2$]Cl,  denoted  
as $\kappa-$ Cl
\cite{Williams90}, is the prototype compound of the series showing  the complete sequence of
states namely, the  Mott-insulating, antiferromagnetic,
  metallic  and superconducting states, within a  
pressure interval of few hundred bars \cite{Ito96,Miyagawa95}. Despite  the numerous experimental
efforts recently expended on the properties of this salt, the information
collected  from experiments done under pressure 
 remained until now scattered and limited by  the selectivity of the
experimental probe used. 
 Regions of
stability of  the metallic and superconducting phases have been
investigated whereas the information about the  pressure profile of  AF critical point 
is missing so far \cite{Ito96,Kanoda97}. Our knowledge on the multicritical
structure of opposing phases and the nature of the MI transition under pressure is also partial    so
that  a major  part of the  phase diagram  remained until now grounded  on a conjectural
rather than  an empirical  basis \cite{Kanoda97,McKenzie97}.     

The experiments that are presented in this work were undertaken in
order to yield
 an accurate phase diagram of $\kappa-$Cl, which is shown in
Figure 1. An 
 hydrostatic helium gas pressure 
technique has been used in order to cover the $P-T$
phase diagram  from  both isothermal and isobar sweeps.  $^1$H NMR and  AC suceptibility techniques
were simultaneously employed and separated sectors of the phase diagram where either AF or SC state   is
stable have been unraveled.
Both phases meet at $(P^*,T^*)\simeq$~(282 bar, 13.1K),  a
point  that ends a first order AF-SC  boundary, which in  turn
separates two  regions  of inhomogeneous 
 coexistence of  AF and SC phases. 
As pressure is swept in the
high-temperature paramagnetic domain, one crosses a first-order 
line  associated with the Mott  transition and which evolves towards a crossover 
above a  critical point where the MI transition line ends.

\begin{figure}[htb]
\epsfxsize 7.5 cm
\centerline{\epsfbox{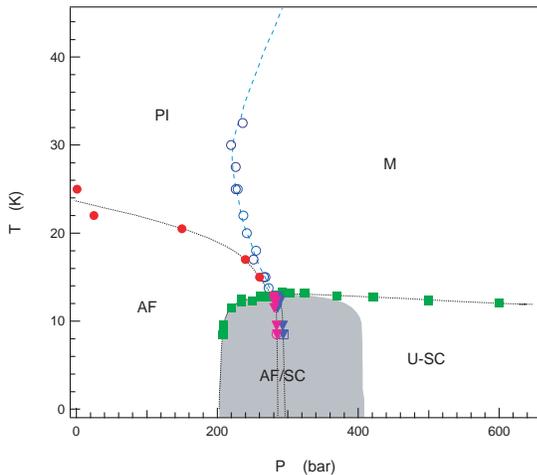}}
\label{}
\caption{Temperature {\it vs} pressure phase diagram of $\kappa-$Cl. The
antiferromagnetic (AF) critical  line $T_N(P)$ (dark circles) was
determined from  NMR relaxation rate while   $T_c(P)$  for  unconventional
superconductivity (U-SC: squares) and the metal-insulator
$T_{MI}(P)$ (MI: open circles) lines were  obtained from the  AC
susceptibility. The  AF-SC boundary (double dashed line) is determined
from the inflexion point  of $\chi'$(P) and, for 8.5K, from  sublattice
magnetization. This boundary line separates two regions
of inhomogeneous phase coexistence (shaded area).}
\end{figure}

All measurements under
pressure were performed on single crystals  of approximately 
0.85~x~0.75~x~0.075~mm$^3$,  synthesized and grown by standard electrochemical
methods
\cite{Williams90}. 
Temperature and pressure sweeps were 
sufficiently slow to show no dependance on the sweeping rate
($\sim 0.07$~K/min,~$\sim 1$~bar/min). In order to ensure  hydrostatic pressure conditions, our
measurements are restricted to the region above the Helium solidification line.  $^1$H NMR measurements
were done in a low static field of 0.5~T oriented perpendicularly to the conducting layers, which
essentially allows to obtain a zero field phase diagram. 
For AC susceptibility, we used a 
zero-phase  feedback loop to track the rf
resonance frequency of the NMR circuit with a sensitivity of few  ppm.
No static field was applied and
the AC magnetic field was  approximately 0.05~Oe in a
direction parallel to  the layers $-$ in order to also statisfy NMR
requirements.

Different field and temperature conditions, namely  1~T and 3~K in the vortex
$-$lock-in $-$ orientation,  have been applied at about 300 bar in a
clamp pressure cell  in an attempt to detect the presence of AF vortex
cores and to establish
the stability of the coexistence region
down to 3~K.
%figure 2

\begin{figure}[htb]
\epsfxsize 8.5 cm
\centerline{\epsfbox{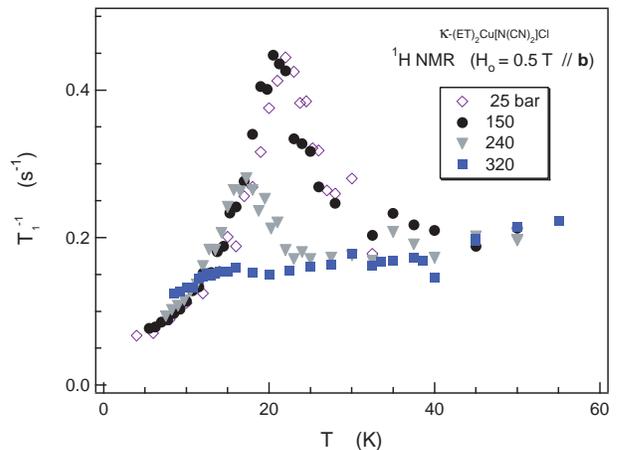}}
\label{t1vst_p}
\caption{Temperature dependence of
$^1$H NMR relaxation rate for $\kappa$-(ET)$_{2}$Cu[N(CN)$_{2}$]Cl at
various pressures. The
static magnetic field (0.5T) was applied perpendicularly to the layers.}
\end{figure}

The results
for the temperature
dependence of  the nuclear relaxation rate $T^{-1}_1$   are shown in Figure
2 for various
pressures.  The singular peak of $T^{-1}_1$,  which marks the onset of  critical
AF ordering at
the N\'eel temperature $T_N$ is gradually suppressed under pressure with the value $dT_N/dP\simeq
-0.025$K/bar   for the pressure coefficient. Above  275 bar, there is no peak
 indicating the absence of a magnetic transition. A plateau
of $T^{-1}_1 $, however, persists up to
40~K or so,  which marks
a sizeable enhancement due to short range AF correlations.

\begin{figure}[htb]
\epsfxsize=0.95\hsize
\centerline{\epsfbox{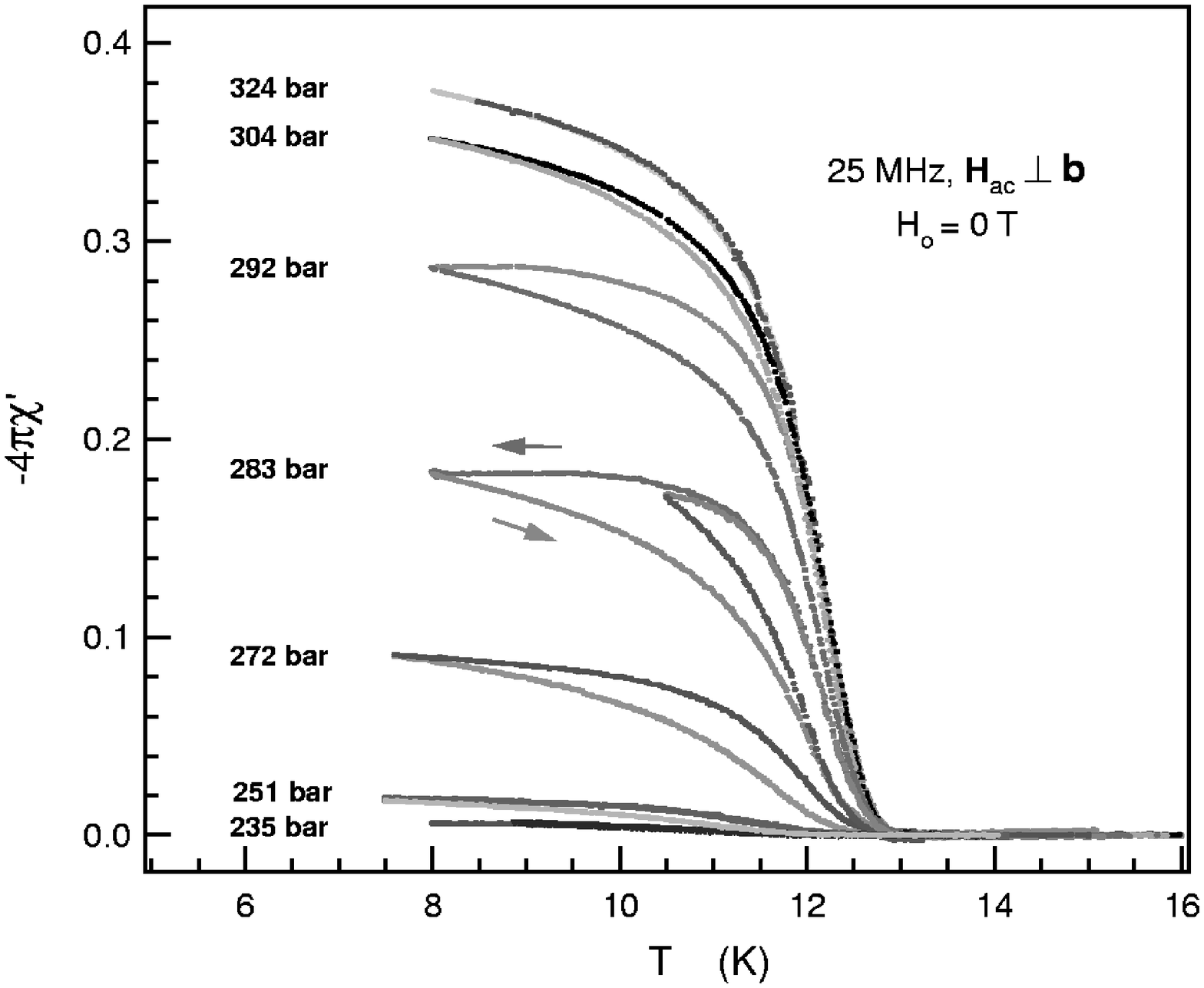}}
\epsfxsize=0.95\hsize
\centerline{\epsfbox{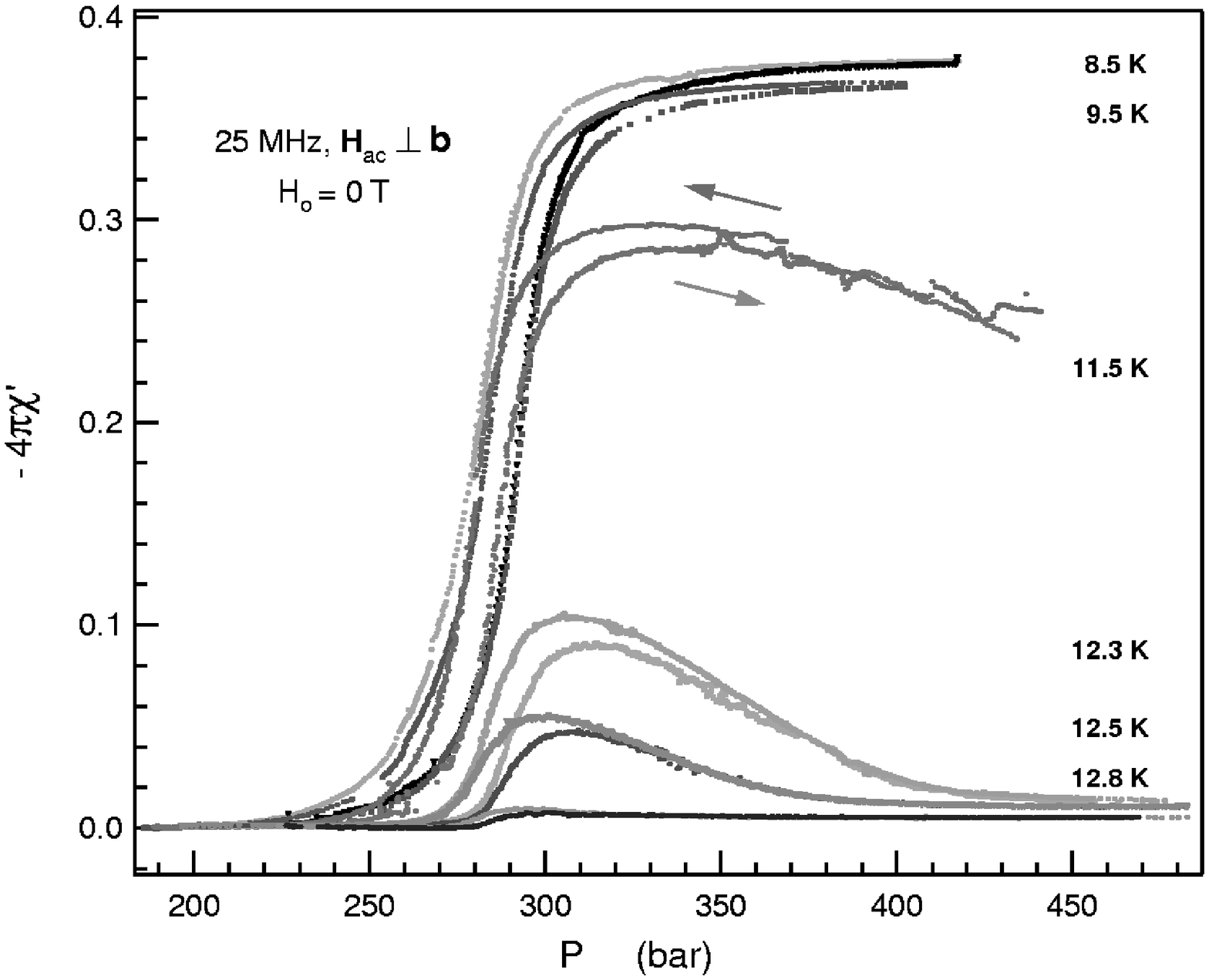}}
\label{}
\caption{Real part of AC susceptibility of $\kappa$-Cl for  a) temperature
sweeps at selected
pressures and b) for  pressure sweeps
at selected temperatures.}
\end{figure}

An accurate determination of superconducting order  in the phase diagram
can be  obtained from the AC
susceptibility  measurements as
shown  in Figure 3 for selected  temperature and
pressure sweeps respectively. In the high pressure domain above 400 bar,  
the $T_c(P)$ line below which there is a finite density of 
superconducting condensate slowly decreases in agreement with previous results ($dT_c/dP \simeq$~ -3.8
K/kbar, squares in Figure 1)\cite{Ito96}. 
As pressure is reduced below 400 bar,  lower
saturation levels are recorded indicating a gradual suppression of the
superconducting order
down to 200 bar where it vanishes;
$T_c(P)$ thus crosses
$T_N(P)$ and  the regions of stability for SC and  AF    overlap.

When the variation of the superconducting condensate is analyzed  under
pressure  between 8.5K
and    12.8~K (Figure 3, bottom), there is a rapid change of the
diamagnetic signal centered around a transition  pressure
denoted      $P_1$ $-$ defined at the inflexion point. At  8.5~K,
$P_1\simeq 290$~bar, and $P_1$ slightly moves
downward   as temperature is raised to finally reach
$P^*$ at $T^*$ (double dashed line in Figure 1). According to Figure 3,  an
hysterisis can be found in a given interval of pressure and
temperature. This region of coexistence  which conveys  some metastability of the transition is delimited
by the shaded area in Fig.1.

The nature of this region of the phase diagram can be further sharpened by
first looking at the
pressure dependence of the $^1$H NMR line shape as shown in Figure 4 for
8.5~K. In the AF phase at
$P=15$~bar, the signal is split into a number of  discrete peaks, a
characteristic of commensurate AF order \cite{Miyagawa95}; this structure
persists up to
200~bar, above which  a narrow peak close to the origin grows in
importance concomitant with  a reduction of the AF line shape structures.
At high pressure, where the system is completely in the SC state, this
peak  dominates the spectrum. A simulation of the line shape  then
allows  to determine the fraction of the sample that becomes either AF or
SC. From Figure 4 (right inset), there is a gradual suppression of  AF order
which becomes  steepest at $P_1\simeq $290~bar and is tied to a concomitant
increase of the SC
order. The 10 bar hysteresis at $P_1$ is  in accordance with the AC susceptibility measurements of
Figure 3.

\begin{figure}[htb]
\epsfxsize=1.00\hsize
\centerline{\epsfbox{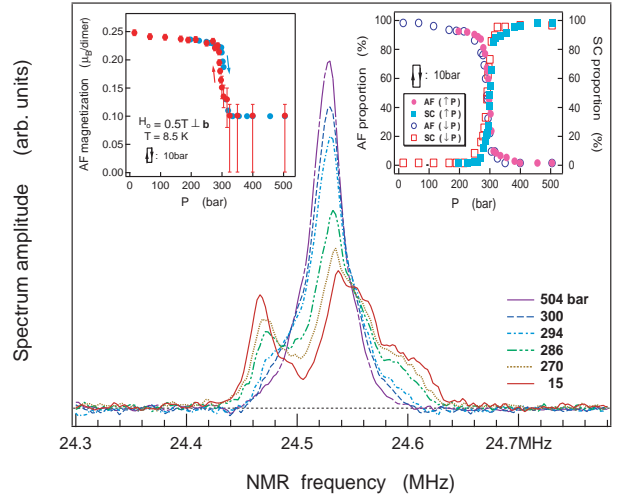}}
\label{spectre}
\caption{  $^1$H NMR spectra of $\kappa-$Cl for  different pressures at
8.5~K. Insets: Pressure
dependence of AF magnetization (left) and  proportion of nuclear spins  in the AF and SC states (right)  as
extracted from the spectrum.}
\end{figure}

%\begin{figure}[htb]
%\epsfxsize=0.95\hsize
%\centerline{\epsfbox{propvspb.eps}}
%\label{}
%\caption{ Pressure dependence of the proportion of nuclear spins found in
%the AF (left) and the
%SC (right) regions at 8.5K.}
%\end{figure}

A further, more direct,  confirmation of the first order
character of the AF-SC transition line is provided by the pressure
variation of the AF order parameter  as
obtained from the NMR spectral line shape. This is exhibited  in the left inset of Figure 4
where
the AF order parameter at
$T=8.5$~K shows a slow decrease  under pressure  followed by
an abrupt drop at $P_1$=290 bar.

At this point, a few remarks are in order.   Although our results do show
evidence of coexisting
phases, they do not allow to determine if a  macroscopic or mesoscopic
(e.g stripes,
\cite{Kivelson94}) type of coexistence is taking place. Given here the constant band filling
(half-filling) 
   under pressure,  the stripe
formation, if it exists, would be of different nature than for high-T$_c$ materials\cite{Kivelson94}.  The
nature of the  point
$(P^*,T^*)$ in the phase diagram  is also of interest. Since it exhibits
hysteresis
(Figs. 3), it can hardly be  classified as a bicritical point, which
is second order in
character.     As we will see shortly, however,  the point $(P^*,T^*)$ also
belongs to another
transition  line associated with the MI transition between two unbroken
symmetry states that is,
the  Mott insulating and the metallic states.

We have
used the  NMR spectra
to check if superconducting vortices with AF cores may be  part  of the SC-rich
sector $-$ AF vortices are predicted in the  SO(5) scenario for unification of magnetism and
superconductivity 
\cite{Zhang97}.
  By looking at the modification of internal field
distribution in NMR
line shape as function of the static field orientation  at $P\simeq 300$~bar and down to 3K, however, 
we failed to  detect any
decrease of the AF part of the spectra when the static field is oriented
along  the
so-called `lock-in' direction. This  orientation corresponds to the situation of
intrinsic pinning of vortices between the conducting planes and  where
vortex cores  should  sustain a sizeable reduction of their magnetic
component. Our results  then indicate  that  the
vortex cores are non magnetic,   at the  very least in this
region of the phase diagram.

 When AC susceptibility measurements are  performed under pressure in
the paramagnetic
temperature domain, a jump in the diamagnetic signal, albeit small in
amplitude,   is
clearly found (Figure 5). It reveals
an increase of diamagnetism  when the system  enters in the metallic phase
$-$ through skin depth
effect. These observations
corroborate    previous  electrical transport measurements by Ito {\it et
al.,} who located  the MI transition in the same pressure and
temperature domain
\cite{Ito96}.
\begin{figure}[htb]
\epsfxsize=0.95\hsize
\centerline{\epsfbox{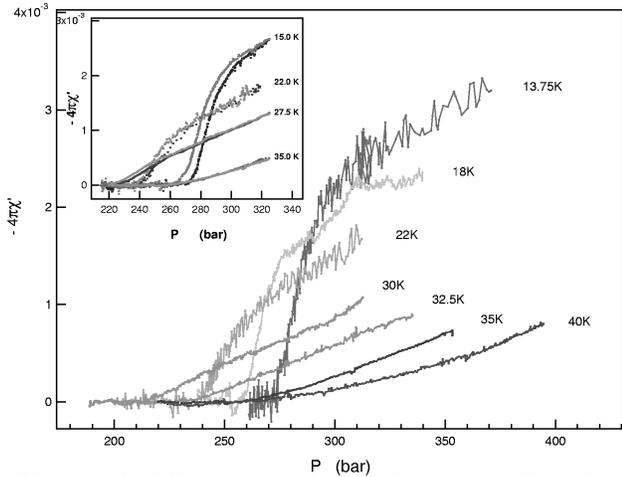}}
\label{traces}
\caption{The MI transition seen by the pressure dependence of AC
susceptibility in the paramagnetic
phase. The inset shows  the variation of hysteresis effects at different
temperatures
.}
\end{figure}
When the temperature is increased, the  jump in the diamagnetic susceptibility
evolves towards a smooth concave profile above some  point $(P_0,T_0)
\simeq (220~{\rm
bar},32.5~{\rm K} )$. Well below this point, the diamagnetic susceptibility
shows a small but
detectable hysteresis  that decreases in amplitude as the temperature is
raised  from
($P^*,T^*)$,  indicating that the MI  transition is first order in
character (inset of Figure
5).   Within experimental
accuracy, the MI line   also starts from  ($P^*, T^*)$ where all other phases
meet\cite{Note1}. The end point
$(P_0,T_0) $ can then be conjectured to be a critical point.

Refering to Figure 1, over most part of the MI-metal equilibrium curve
$dP/dT$ is  negative.
According to the Clausius-Clapeyron  relation a negative
$dP/dT$ would indicate a reduction  of the spin entropy on the insulator
side  below the
metallic level.  A reasonable explanation for this could come from low dimensional short-range
AF correlations, which
   extend relatively deep in the paramagnetic domain (Figure 2)
 and would quench a sizeable part of the spin entropy in
the vicinity
of $T_N(P)$. Sufficently far from the AF transition, however, entropies
nearly balance so that $dP/dT$ is close to zero around 30~K or so\cite{Note2}.

As for global phase diagram, important
conclusions may be drawn about the
description  of this series of layered organic superconductors. It is noticeable in the first place that
the MI transition is discontinuous and clearly
evolves toward a mere crossover above a critical point ($P_0,T_0$). To our
knowledge, it seems to
be the  first time that  a genuine electronic transition that combines all
these characteristics
is  discovered in a quasi-two-dimensional system at half-filling \cite{Georges96}. In the
second place, the
joining of the MI line with
$T_N(P)$ at  ($P^*, T^*)$ is of great interest  since it
shows within experimental accuracy the absence of boundary between the
metallic and
a complete AF phases. This  confirms previous  inferences made about the
absence of
itinerant  antiferromagnetism in $\kappa-$(BEDT-TTF)$_2$X  \cite{Kanoda97}
and the relevance of a  description of magnetic ordering    
 in terms of interacting spins localized on dimers
\cite{McKenzie97,Kino96}.   The
fact that SC and AF phases  overlap  below $P^*$   indicates that
superconductivity  can be  directly stabilized from the insulating phase.
 An inescapable outcome  of this result is the obvious
exclusion of a weak
coupling scenario for  the emergence of unconventional
pairing in layered organic superconductors as function of pressure.

The existence of the point-like region at $(P^*,T^*)$ where metal, Mott
insulator, antiferromagnet and
non s-wave superconductor all meet demonstrates that strong
electron correlations and broken symmetry variables  are equally important
for a unified
theory of antiferromagnetism and non s-wave superconductivity in these
compounds \cite{Note3}.

\noindent
{\bf Acknowledgements}.$-$ The authors  thank K. Behnia, L. G. Caron,  V.
J. Emery, A.
Georges, C. Kallin, S. Kivelson, R. B. Laughlin,  D. S\'en\'echal, A.-M.
Tremblay and D. Zanchi for
useful discussions.  We are grateful to N. Nardone for his  crucial
technical support and precious
advices.  P. W. would like thank the NSF for financial support during his
stay at UCLA . C.B thanks the NSERC and
the Canadian Institute for Advanced Research  for financial support.

\end{document}